\documentclass[referee]{raa}            

\usepackage{graphicx,times}             
\usepackage{natbib}
\usepackage{amssymb,amsmath}
\bibpunct{(}{)}{;}{a}{}{,}

\usepackage[a4paper=true,dvipdfm=true,pagebackref=true]{hyperref}
\hypersetup{colorlinks = true, linkcolor = green, anchorcolor = red, citecolor = blue, filecolor = red, pagecolor = red, urlcolor = red}

\begin{document}
   \title{Spectro-polarimetric Observations at the NVST: I. Instrumental Polarization Calibration and Primary Measurements}

 \volnopage{ {\bf 20XX} Vol.\ {\bf X} No. {\bf XX}, 000--000}
   \setcounter{page}{1}

   \author{Jun-Feng Hou\inst{1}, Zhi Xu\inst{2,*}, Shu Yuan\inst{2}, Yu-Chao Chen\inst{2}, Jian-Guo Peng\inst{2}, Dong-Guang Wang\inst{1}, Jun Xu\inst{2}, Yuan-Yong Deng\inst{1}, Zhen-Yu Jin\inst{2}, Kai-Fan Ji\inst{2}, Zhong Liu\inst{2}
   }
   \institute{ National Astronomical Observatories, Chinese Academy of Sciences, Beijing 100101,
China; \\
        \and
        Yunnan Observatories, Chinese Academy of Sciences, Kunming 650216, China\\
}

\footnotetext{\small $*$ Author for correspondence: e-mail: xuzhi@ynao.ac.cn.}
\abstract{This paper is devoted to the primary spectro-polarimetric observation performed at the New Vacuum Solar Telescope of China since 2017, and our aim is to precisely evaluate the real polarimetric accuracy and sensitivity of this polarimetry by using full Stokes spectro-polarimetric observations of the photospheric line Fe I 532.4 nm. In the work, we briefly describe the salient characteristic of the NVST as a polarimeter in technology and then characterize its instrumental polarization based on the operation in 2017 and 2019. It is verified that the calibration method making use of the instrumental polarization calibration unit (ICU) is stable and credible. The calibration accuracy can reach up to 3$\times 10^{-3}$ . Based on the scientific observation of the NOAA 12645 on April 5th, 2017, we estimate that the residual cross-talk from Stokes $I$ to Stokes $Q$, $U$ and $V$, after the instrumental polarization calibration, is about 4$\times10^{-3}$ on average, which is consistent with the calibration accuracy and close to the photon noise. The polarimetric sensitivity (i.e., the detection limit) for polarized light is of the order of $10^{-3}$ with an integration time over 20 seconds. Slow modulation rate is indeed an issue for the present system. The present NVST polarimeter is expected to be integrated with an high-order adaptive optics system and a field scanner to realize 2D magnetic field vector measurements in the following instrumentation update.
\keywords{Techniques: polarimetric -techniques: spectroscopic - Sun: magnetic fields}
}

   \authorrunning{Jun-Feng Hou et al. }            
   \titlerunning{Instrumental Polarization Calibration using the Spectro-polarimetric mode of the NVST}  
   \maketitle

%
\section{Introduction}           
\label{sect:intro}

The 1-meter New Vacuum Solar Telescope (NVST) located at Fuxian Solar Observatory aims to observe dynamic solar structures and vector magnetic fields with high resolution \citep{LiuZ2014}. To achieve that, its terminals comprise a multi-channel high-resolution imaging system \citep{XuIAU2014} and a high-dispersion multi-line spectrograph \citep{WangR2013}. Since 2017, the spectrograph has been improved to develop the capability of spectro-polarimetric observation. The spectro-polarimetric observation can be carried out in one or several visible spectral lines. Full Stokes profiles are all obtained, from which the vector magnetic field can be retrieved by resolving the radiative transfer for polarized radiation with an atmospheric model assumption and applying many sophisticated inversion technique \citep{Harvey1972,Skumanich1987,Lagg2004}

There is a common problem that the polarimetric data measured by the polarimeter needs to be calibrated for the cross-talk (i.e., the linear combination) between different polarization states caused by the polarimeter itself \citep{delToroIniesta}. It can be described as $S_{obs} = X \cdot S_{sun}$, where $S_{sun}$ is the incident Stokes vector from the Sun to the telescope. $S_{obs}$ is the direct data product from the observation. $X$ represents the response matrix (function). The measurement accuracy of polarimetry largely depends on the accurate knowledge of the matrix $X$. Besides many efforts to precisely build the theoretical polarization model of the entire system, since 1980s the knowledge of $X$ can be experimentally determined in aid of a device, which can produce an incident light with a known polarization state \citep{Baur1980}. It is usually called the instrumental polarization calibration unit (ICU or PCU). This technique has been utilized by several telescopes, such as the Advance Stokes Polarimeter at vacuum tower telescope of NSO \citep{Skumanich1997}, the Domeless Solar Telescope at Hida observatory \citep{Kiyohara2004, Anan2018}, the German Vacuum Tower Telescope on Tenerie \citep{Beck2005}, the Solar Optical Telescope aboard Hinode \citep{Ichimoto2008} and the 1.5 m solar telescope GREGOR \citep{Hofmann2012} and so on. An ICU designed by Huairou Solar Observing Station has been equipped on the NVST since 2017 \citep{Hou2017} in order to precisely determine the matrix $X$.

In this work, we report the current spectro-polarimetric observation performed at the NVST and figure out the main polarization characteristic of the NVST, including the polarization modulation mode, the procedure to calibrate and remove the instrument polarization, the actual polarimetric accuracy and sensitivity, etc. These analysis are mainly based on the spectro-polarimetric observations run in 2017 and 2019. It is worth noting that, both the present observation and analysis are preliminary steps in view of the general NVST objectives. Especially the measurement procedure is a preliminary step for high precision polarimetry observation since we presently adopt a temporal and single-beam step-wise modulation. We have ignored the measurement error caused by the atmosphere seeing and expect that the spatial smearing can be mitigated by using the adaptive optics (AO) technique. One high-order solar AO system has been developed \citep{Rao2016}, and combination with the AO system is our following instrumentation update.

In Section 2, we describe the main characteristics of the spectro-polarimetric observation system. In section 3, we present our polarization modulation method and emphasize the importance of the calibration of the instrument polarization to our polarimetry. Section 4 deals with the calibration method and procedure. We estimate the calibration accuracy and the influence of the spectrograph position angle using the continuous calibration carried out in 2017 and 2019. Scientific observations for an active region are presented in Section 5 used to verify the calibration credibility and evaluate the real polarimetric sensitivity of our polarimetry.

\section{Salient features of the Spetro-polarimetric observation system at the NVST}

Figure 1 presents the optical system of the NVST polarimeter, mainly including a 98-cm aperture vacuum telescope alt-azi mounted, a vertical multi-wavelength spectrograph, polarization modulator.

An ICU consisting of a linear polarizer and a $\lambda/4$ retarder is installed at the secondary focus (F2). As demonstrated in the upper-left corner of Figure 1, it is mounted in the front of a cylindrical device, which has a central hole in the radial axis and allows the light to go through along the optical axis for scientific observations. When we perform the calibration, the device rotates and its cylinder axis is moved to the optical axis direction, which meanwhile enables ICU to enter the optical path \citep{Qin2018}. The effective aperture of ICU is about 36 mm. We utilize a Wire Grid Versalight polarizer from Meadowlark Optics. This kind of polarizer offers the broadest and highest field of view and reflects another polarization state, which can decrease the thermal expansion effect in vacuum system. The retarder used is a zero-order achromatic waveplate consisting of quartz and MgF2 materials. It is manufactured by Union Optics. The thicknesses of quartz and MgF2 materials are 0.521 mm and 0.415 mm, respectively, which make sure the retardation variation within 90$^{\circ}$ $\pm$ 10$^{\circ}$ in a wide range of wavelength from 520 nm to 1083 nm. Retardance measurements at several wavelengthes of interest are given in Table 1. In order to improve the parallelism of the retarder, a double-separation structure is adopted, which makes the beam-deviation of each plate (quartz or MgF2) less than 1 arc second. The reflectivity is less than 0.1 \% by coating anti-reflective films. Both the polarizer and retarder can be rotated independently relative to the reference axis (i.e., the axis of the polarizer at its home position), which is parallel to the telescope elevation and defines the direction of the Stokes $+Q$ in the analysis. Requirements on the accuracy of the polarization axis and the gear control of the stepping motor are about 0.5 $^{o}$ and 0.002$^{o}$. Because there is no short passway between ICU and the observation room, we have to remotely control ICU by using the wireless technique.

\begin{table}
\begin{center}
\caption[]{Retardance of the ICU waveplate at different wavelengthes }\label{Tab:1}
{\begin{tabular}{|c|c|}
\hline
 wavelength (nm)  & retardance (degree)  \\
\hline
 525.0  & 90.72 \\
 \hline
 532.4  & 91.74 \\
 \hline
 617.3  & 98.54 \\
 \hline
 854.2  & 92.81 \\
 \hline
 864.8  & 92.08 \\
 \hline
 1074.7  & 84.38 \\
 \hline
 1083.0  & 85.04 \\
 \hline
\end{tabular}}
\end{center}
\end{table}

In the Coud\'{e} laboratory as shown on the right of Figure 1, the main terminals are composed of a high resolution imaging system and a high dispersion spectrograph. They are arranged perpendicularly to each other. The entrance slit of the spectrograph is just set at the Coud\'{e} focus of the telescope (F3). It is noted that in the present, both the AO system and 2D field scanner (enclosed by a dotted box) have not been co-operative with the spectro-polarimetric observation and ignored from the current work. Before the focus F3, we utilize a beam splitter ($M_{45}$) to distribute the incoming light into the imagining system and spectrograph. In this situation, slit-jaw images can be reordered by using the imaging system. All the backends in the laboratory are relatively static, but they can be synchronously rotated as a whole  in the azimuth orientation of the spectrograph in order to compensates the image rotation.

The multi-wavelength spectrograph can measure three spectrum lines simultaneously, including two chromospheric lines (H$\alpha$ and Ca II 854.2 nm) and one photospheric Fe I lines at 532.4 nm. In order to concentrate on the spectro-polarimetric observation, $M_{45}$ is specially designed to allow a wavelength band of 532.4 $\pm$ 5 nm to enter the spectrograph and reflects the rest light into the imaging system monitoring the slit position. As a result, H$\alpha$ and Ca II spectrum observations are not available. Instead, high resolution imaging observations at some specific wavelengthes (e.g. the H$\alpha$, TiO-band or G-band) can be obtained simultaneously with the spectro-polarimetric observation.

The Fe I 532.4 nm line is sensitive to the magnetic field with a Lander factor $g$ of 1.5. It is also employed by the Solar Magnetism and Active Telescope (SMAT) at Huairou Solar Observing Station \citep{Ai1986}. A 1200 mm$^{-1}$ grating is used to get the Fe 532.4 nm line from the 2nd order onto the camera. The camera is a 14-bit PCO4K CCD with a full size of 4008$\times$2672 pixels, and a pixel size is of 9$\mu$$\times$9$\mu$. The saved data are 2-pixel binning over columns and rows in order to increase the signal level. The wavelength sampling is of 0.001 nm pixel$^{-1}$ after binning and wavelength range is about 1.336 nm (i.e., 0.001$\times$ 2672 /2. = 1.336 nm). By comparison with the solar spectral atlases provided by the Fourier Transform Spectrometer (FTS) at the McMath/Pierce Solar Telescope, we can estimate the instrumental profile (a Gauss profile with a certain full-width-of-half-maximum, FWHM) and the stray light ($\alpha$) by minimizing the $\chi^{2}$ value, expressed as
\begin{equation}
\chi^{2}(FWHM, \alpha) = [I^{obs}_{\lambda} - (I^{FTS}_{\lambda}(1-\alpha) \bigotimes \ G(FWHM) + \alpha)]^{2}
\end{equation}

Here the operator $\bigotimes$ represents the convolution and $I_{\lambda}$ is a normalized profile to the continuum. One observed profile compared with the modified FTS profile is shown in Figure 2. The minimum deviation between them can be obtained as the FWHM is equal to 118 $m\AA$ and the $\alpha$ equal to 0.6 $\%$.

An assembly of polarization modulator and analyzer is installed in front of the spectrograph slit. It is composed by a step-wise rotating retarder and a fixed linear polarizer. Both are manufactured by Huairou Solar Observing Station. Both the modulator and the analyzer have an effective apertures of 36 mm. The retardance of the modulator is optimized to be 127$^{\circ}$ at 532.4 nm to obtain the equal modulation efficiency for different Stokes states. It is measured to be 127.8$^{\circ}\pm 0.02^{\circ}$ in the laboratory. The angle between the fast axis of the modulator at the origin and the transmission axis of the analyzer is accurately measured to be 31.06$^{\circ} \pm 0.06^{\circ}$ in the laboratory and taken into account in the demodulation process. Next, the transmission axis of the analyzer is placed parallel to the slit orientation. Inevitably, there may be an angle offset due to the installation error. Since it is not so easy to precisely measure this angle offset, we prefer to use a mechanical positioning to fix the assembly position with respect to the slit and ensure the reset accuracy. In addition, some interference fringes are apparently present in the spectrum data. In order to efficiently eliminate the fringes, we tilt the assembly to make a small wedge angle between the glass surface and the slit surface.

\section{Spectro-polarimetric observation of the NVST}
In the present NVST configuration, the polarization modulation is performed by step-wise rotations of the retarder and the intensity of only single beam selected by the analyzer is measured. The modulated beam intensity ($I_{obs}$) as a function of four incoming Stokes parameters ($S_{in}$), the modulator rotation angle ($\theta$) and its retardance ($\delta$) is given by:
\begin{equation}
I_{obs}(\theta, \sigma) = I_{in} +Q_{in}(cos^22\theta + sin^{2}2\theta cos\delta) + U_{in}[sin2\theta cos2\theta(1-cos\delta)] - V_{in}(sin2\theta sin\delta )
\end{equation}
or expressed using matrices,
\begin{equation}
I_{obs} =  M \cdot S_{in}
\end{equation}
Here $S_{in}$ means the incoming Stokes parameters \textit{toward} the modulator. $M$ represents the modulation process of the retarder. In order to derive the full Stokes parameters, the retarder is rotated to 8 positions step-by-step with an interval of 22.5$^{\circ}$ in one half rotation (i.e., $\theta =0^{\circ}, 22.5^{\circ}, ..., 180^{\circ}$). The gear control of the stepping motor is of 0.002$^{\circ}$. CCD takes images synchronously with the positioning of the retarder. To reach high polarization sensitivity we have to recorder $n$ frames at each position. The camera does not record any images in the second half rotation between 180$^{\circ}$ to 360$^{\circ}$. As a result, we construct an 8$\times$4 modulation matrix $M$ and obtain an observed $8 \times 1$ intensity matrix $I_{obs}$, i.e., [$I^{1}_{obs}$, $I^{2}_{obs}$, ... $I^{8}_{obs}$]$^{T}$. Here $I_{obs}^{i}$ represents the spectral image integrated over $n$ frames.

An appropriate demodulation (i.e., successive addition and substraction of images) is then applied, which is represented by a demodulation matrix $D$. The process may be expressed as

\begin{equation}
\begin{split}
D \cdot I_{obs} &= D \cdot M \cdot S_{in}  \\
                &= D \cdot M \cdot M_{M_{45}} \cdot M_{T} \cdot S_{sun} \\
S_{out} &= X \cdot S_{sun}
\end{split}
\end{equation}
Here $S_{sun}$ means the Stokes parameters generated by the Sun itself. $S_{out}$ is the measured Stokes parameters demodulated from the observed intensities. $M_{T}$ means the Mueller matrix of the telescope, followed the Mueller matrix of the beam splitter, $M_{M_{45}}$, along the optical train. In other words, we have $S_{in} = M_{M_{45}} \cdot M_{T} \cdot S_{sun}$ in the case of the scientific observation. We also can define a response matrix expressed as $X = D \cdot M \cdot M_{M_{45}} \cdot M_{T}$ and deduce the finial expression in Eq.(4). Only if $X$ is precisely determined, the solar polarimetric signal $S_{sun}$ can be correctly retrieved. Here the process to determine $X$ is called the calibration of instrumental polarization. Being a alt-azimuth modified Gregorian telescope, the NVST has a strong instrumental polarization, i.e., the response matrix seriously varies with the change of telescope pointing position during a day.

\begin{figure}[htbp]
\includegraphics[width=8.5 cm,angle =0]{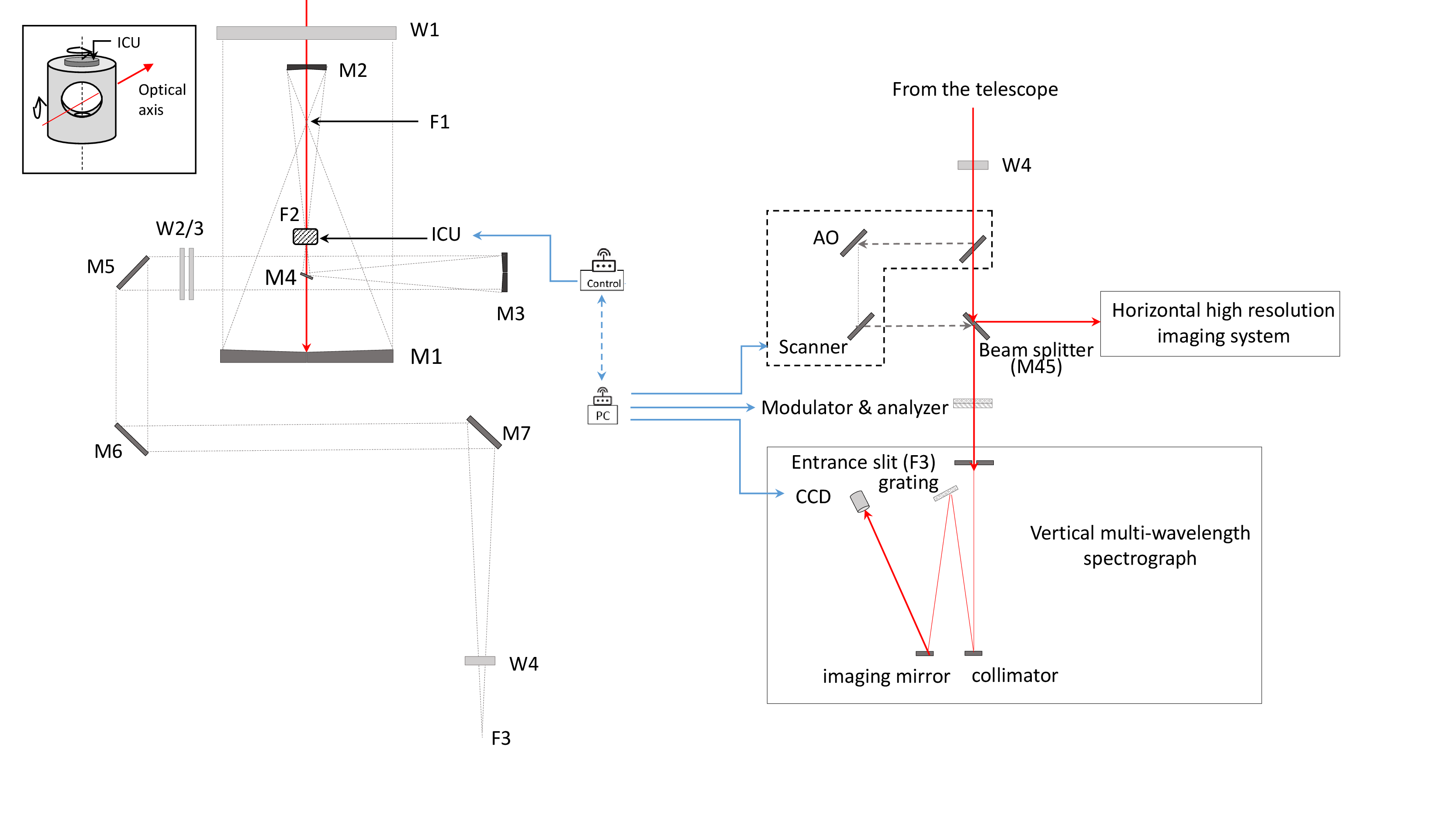}
\caption{Optical layout of the spectro-polarimetric observation system of the NVST. The telescope assembly is shown on the left. The main optics is an on-axis Gregorian including the primary (M1) and secondary (M2) mirrors. M1 has paraboloid figure and M2 has ellipsoid one. After the secondary focus (F2), it is a Coud\'{e} train feeding the light beam down to the laboratory by using a relay mirror (M3, ellipsoid figure) and four folding mirrors (M4, M5, M6 and M7). All optical elements are installed in two evacuated tubes, which are sealed by vacuum windows W1-2 and W3-4, respectively. An ICU is installed at the secondary focus (F2) and its mechanical device is demonstrated at the upper left corner. Backends in the laboratory are shown on the right. Both the AO system and field scanner enclosed by a dotted box are not included in the current optical path, which is indicated by red lines. See the text for more details of the spectrograph, the polarization modulator and analyzer. ICU is remotely controlled though a wireless technique by a PC in the observation room, which also controls the polarization modulator and cameras.}
\label{Fig1}
\end{figure}

\begin{figure}[htbp]
\centering
\includegraphics[width=15.cm,angle =0]{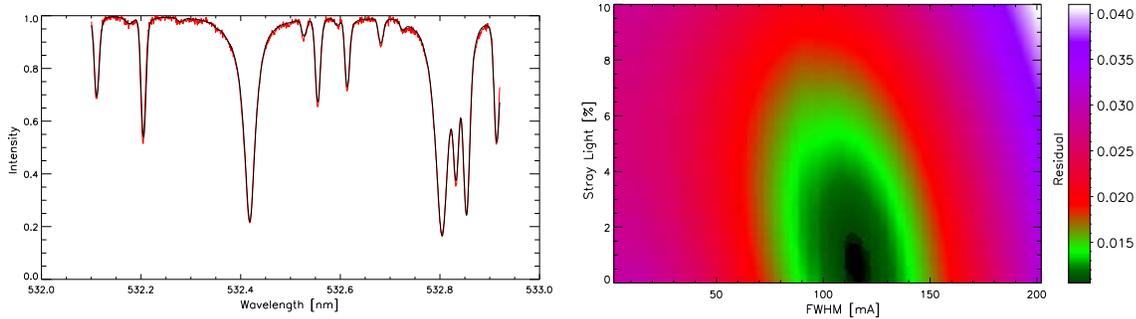}
\caption{Comparison of the observed spectrum profile with the FTS atlas around Fe I 532.4 nm. Left: The FTS profile (the black line) modified by the instrumental profile and stray light is over superposed on the observed profile (the red line). Right: the residual $\chi^{2}$, as defined in Eq.(1), between these two profiles with respect to the changes of the FWHM of the instrument profile and the stray light. }
\label{Fig2}
\end{figure}

\section{Instrumental Polarization Calibration of the NVST }

Based on Eq.(4), we can use a constructed input Stokes vector to replace the unknown $S_{sun}$ taking advantage of ICU in order to derive the matrix $X$ at a specific time. ICU can be moved into the vicinity of F2 if needed, which indicates that we can not take into account the polarimetric properties of the first vacuum window (W1), the primary and secondary mirrors (M1 and M2) that are all located in front of ICU. Here we assume that their polarization effects are negligible since they are axi-symmetric about the optical axis.

\subsection{Measurement of the response matrix $X$}

Firstly, we make the telescope point to the quiet region around the solar disk center assuming the incoming light is taken as unpolarized \citep{Stenflo2005}. Meanwhile the spectrograph, together with the $M_{45}$, the polarization modulator and analyzer, is fixed at the initial position $(\theta_{sp} =0)$ during the calibration process despite of the solar image rotation. Influence of the spectrograph azimuth angle will be discussed in detail in Section 4.3.

Secondly, ICU successively generates 6 kinds of polarized light by rotating its polarizer and retarder independently. The 6 couples of rotation angle ($\theta_{p}$, $\theta_{r}$) are [0$^{\circ}$, 0$^{\circ}$], [45$^{\circ}$, 45$^{\circ}$], [90$^{\circ}$, 90$^{\circ}$], [135$^{\circ}$, 90$^{\circ}$], [135$^{\circ}$, 135$^{\circ}$], [135$^{\circ}$, 180$^{\circ}$], respectively. Thus the constructed input Stokes vector is given by,
\begin{equation}
  S^{c}_{in} = M_{r}(\theta_{r}) \cdot M_{p}(\theta_{p}) \cdot [1, 0, 0, 0 ]^{T}
\end{equation}
Where $M_{p}$ and $M_{r}$ represent the Mueller matrix of the polarizer and retarder in ICU, respectively. The $S^{c}_{in}$ of the 6 times actually make up a 4$\times$6 matrix.

To measure each kind of input Stoke vector, we perform a 8-step modulation spectro-polarimetric observation as described above. At each modulation step we intend to take 5 frames continuously. Meanwhile, more pixels binning and shorter exposure time are adopted to recorder a single frame. After adding 5 frames taken at one modulation state, we further integrate the continuum spectrum both in the wavelength direction (about 0.1 nm) and the spatial direction (about 1/5 of the slit length) to get the observed intensity value $I_{obs}^{c}$,  which is a wavelength- and spatial-independent value. As a result, an 8$\times$6 matrix is composed by the measured intensities, and it can be converted to a 4$\times$6 matrix after demodulation (i.e., $S_{out}$). We show both the constructed $S^{c}_{in}$ and measured (or demodulated) $S_{out}$ in Figure 3.

Eventually, The response matrix $X$ can be derived from the following equation by matrix inversion method.
\begin{equation}
S_{out} = X \cdot S_{in}^{c}
\end{equation}

It totally takes about 53 seconds to finish one complete calibration measurement, which indicates that we assume the polarization property of the polarimeter is stable during this period.

\begin{figure}[htbp]
\centering
\includegraphics[width=12.cm,angle =0]{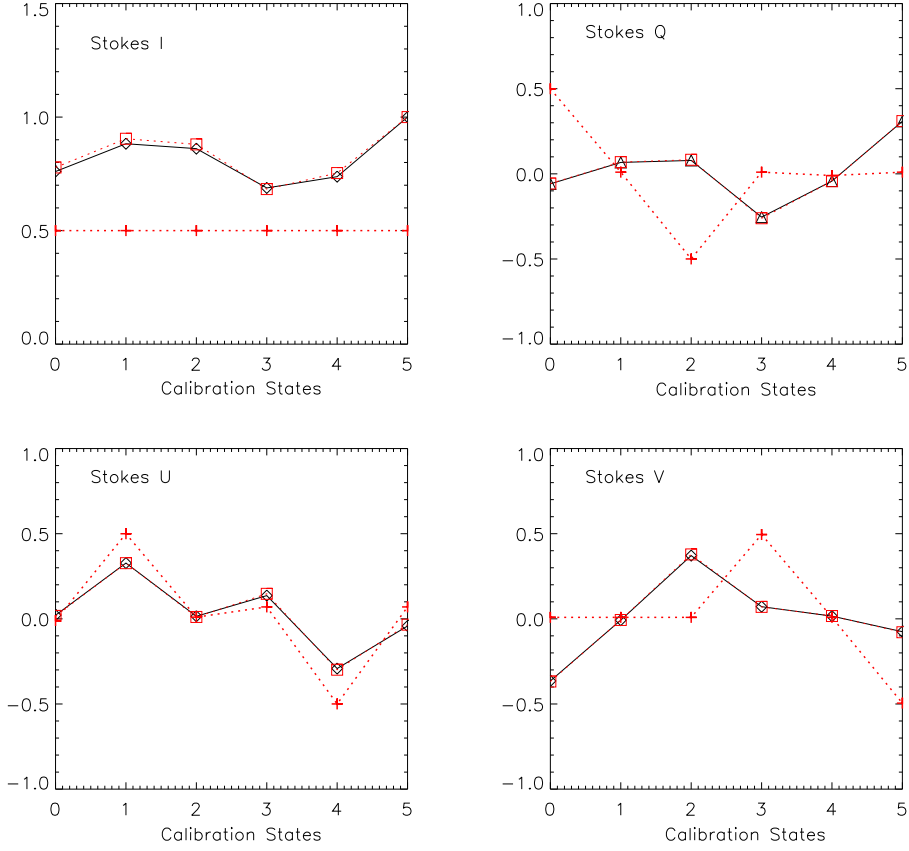}
\caption{Stokes vectors ($I, Q, U, V$) of a calibration measurement. \emph{Red Dash-crossed }: $S_{in}^{c}$, the constructed input Stokes vectors of 6 times, which are calculated by a given retardance and 6 pairs of rotation angles of ICU. \emph{Black Solid-diamonded}: $S_{out}$, the measured Stokes vectors, which are demodulated from the observed intensities. \emph{Red Dot-squared}: the predicated output vectors, i.e., the term of $X \cdot S_{in}^{c}$.}
\label{Fig3}
\end{figure}

\subsection{Calibration accuracy}
The calibration accuracy strongly depends on our knowledge about the $S_{in}^{c}$, or particularly speaking, the polarization properties of the optical elements of ICU. Therefore in addition to the 16 elements of the response matrix $X$, two more free parameters related to ICU are taken into account as we construct the $S_{in}^{c}$.

\begin{itemize}
  \item The retardance of the ICU retarder ($\delta_{r}^{c}$). The value tested in the laboratory is regarded as an initial value.
  \item An offset angle ($\theta^{c}_{off}$) between the ICU polarizer and the retarder axis account for the misalignment during the installation.
\end{itemize}

A change of $\delta_{r}^{c}$ and $\theta_{off}^{c}$ only affects the input vectors $S_{in}^{c}$. All these 18 free parameters can be instantly calculated by minimizing the residual with respect to these two parameters by a gradient method \citep{Beck2005}, as expressed
\begin{equation}
\chi^{2}_{i}(\delta^{c}_{r}, \theta^{c}_{off}) = \sum_{i=0...3}(X \cdot S_{in}^{c}(\delta^{c}_{r}, \theta^{c}_{off}) - S_{out})^{2}_{i}
\end{equation}
$S_{out}$ is the directly measured output vectors, and the term of ($X \cdot S_{in}^{c}$) can be regarded as the predicated output. Once the $\chi^{2}$ reaches the minimum, the solution of $\delta_{r}^{c}$ and $\theta^{c}_{off}$ are determined. The $X$ is finally obtained.

\begin{figure}[htbp]
\centering
\includegraphics[width=12.cm,angle =0]{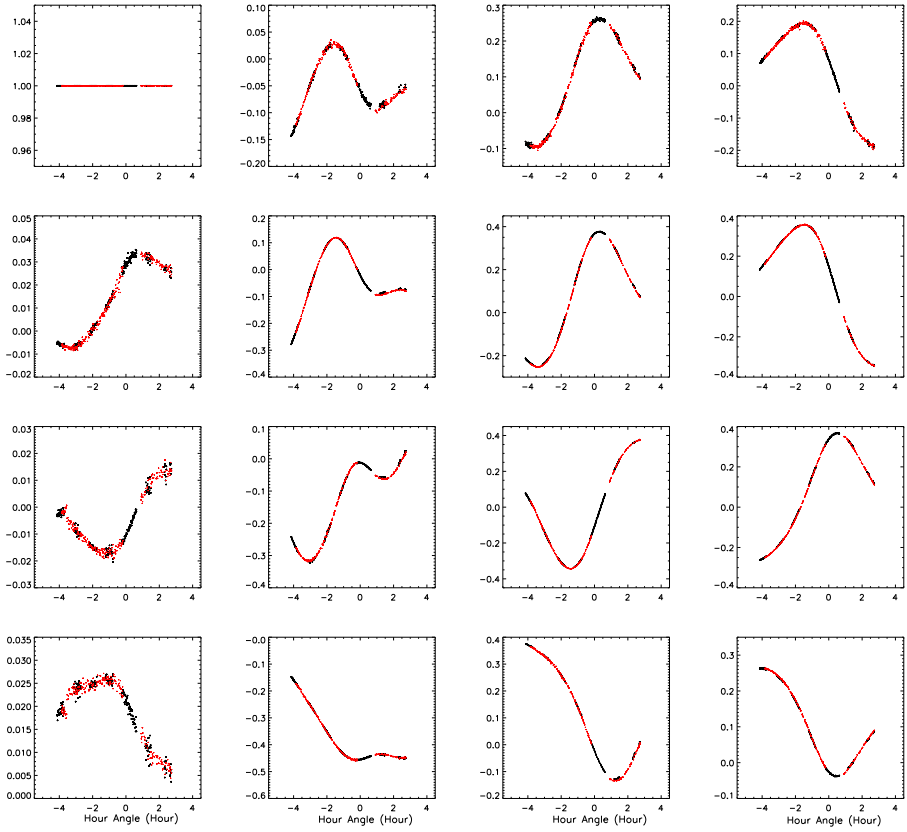}
\caption{Response matrix $X$ of NVST as function of hour angle of the Sun in about 7 hours. It consists of the measurements on February 15th (black dots) and 16th (red dots), 2017. }
\label{Fig4}
\end{figure}

In Figure 3, it is clearly seen that the directly measured vector $S_{out}$ deviates far from the input Stokes vector $S_{in}^{c}$, but shows a good agreement with the predicated output vector. It means that we can well reproduce the output vector from $S^{c}_{in}$  with the correct response matrix $X$. In other words, we can retrieve the $S_{in}^{c}$ from the observed vector as long as the response matrix $X$ is correctly determined. The largest deviation is present in Stokes $I$, which is consistent with other authors reports \citep{Beck2005}. Consequently, we estimate the calibration accuracy from the deviation, i.e., the residual value between the measured and the final predicated $S_{out}$, expressed as $\sum_{n}|X \cdot S_{in}^{c} - S_{out}|$ \ \ (n = 0...5) for each Stokes vector.

In order to investigate the time evolution of the response matrix $X$ during a day and the calibration accuracy and stability, we carry out several times of one-day continuous calibration in 2017 and 2019 (It is reminded that the spectrograph is not available in 2018).

In Figure 4, we show a typical response matrix measured from a 7-hour continuous calibration. It actually comprises the measurement on February 15th and 16th, 2017. There is no apparent discontinuity between these two successive days and all the matrix elements smoothly change with time. Elements in the first column of the matrix, which indicate the cross-talk from $I$ to $I$, $Q$, $U$ and $V$, respectively, are more diffuse at first glance, but their variations are actually one-magnitude lower than others.

In Figure 5, we present the time evolution of the residual value between the measured and the predicated output vectors (Stokes $Q$, $U$ and $V$, respectively) during this period. The averaged residual is about 3$\times$10$^{-3}$ and the standard deviation is about $ 1.3\times10^{-3}$. In detail, from the results of the whole day, it is seen that the residuals of Stoke $Q$ and $U$ are larger than that of Stokes $V$, particularly after hour angle of -1$^{h}$. The largest residual ($\sim 6.6\times10^{-3}$) is found in Stokes $U$ at hour angle of about 2$^{h}$. In addition, the Stokes $Q$ residual after -1$^{h}$ even shows a fluctuation in a sinusoidal distribution, while that of Stokes $U$ shows a negative sinusoidal distribution. It is implied that they are closely related.

\begin{figure}[htbp]
\centering
\includegraphics[width=12.cm,angle =0]{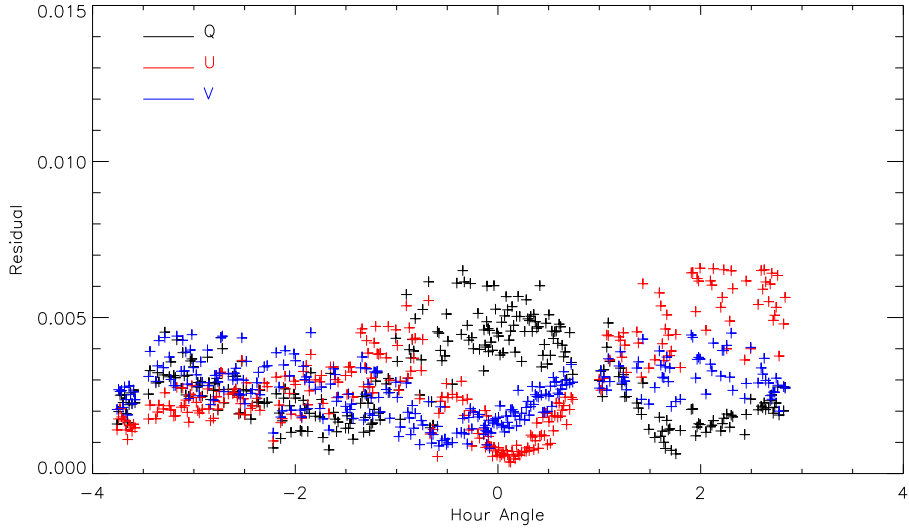}
\caption{Time evolutions of the residual between the measured and calculated Stokes vectors using the correct response matrix $X$. Stokes $Q$ is shown in black, $U$ in red and $V$ in blue. The observation data is the same as the one in Figure 4.}
\label{Fig5}
\end{figure}

As we mentioned above, the retardance of the ICU retarder ($\delta_{r}^{c}$) is considered to be a free parameter during the calculation. It is further revealed that this retardance regularly changes during the continuous calibration and shows a correlation with the variation of the Stokes $I$ intensity obtained by the demodulation. As displayed in Figure 6, the intensity of Stokes $I$ shows an increase-stable-decrease trend from hour angle of -4$^{h}$ to 3$^{h}$, which is closely related to the change of solar elevation angle. During this period, $\delta_{r}^{c}$ gradually increases from $92^{o}$ to $98^{o}$ and then becomes stable around the culmination time, but there is no obvious decrease since then. It is worth noting that the retardance after hour angle of -1$^{h}$ shows a large diffusion range than before.

In contrast, another free parameter, the off-set angle ($\theta^{c}_{off}$) has a steady value of around 0.35 degree. Since this value is fairly stable not only on February 15th and 16th but also on other days. We eventually fix it and take into account the left 17 parameters.

\begin{figure}[htbp]
\centering
\includegraphics[width=12.cm,angle =0]{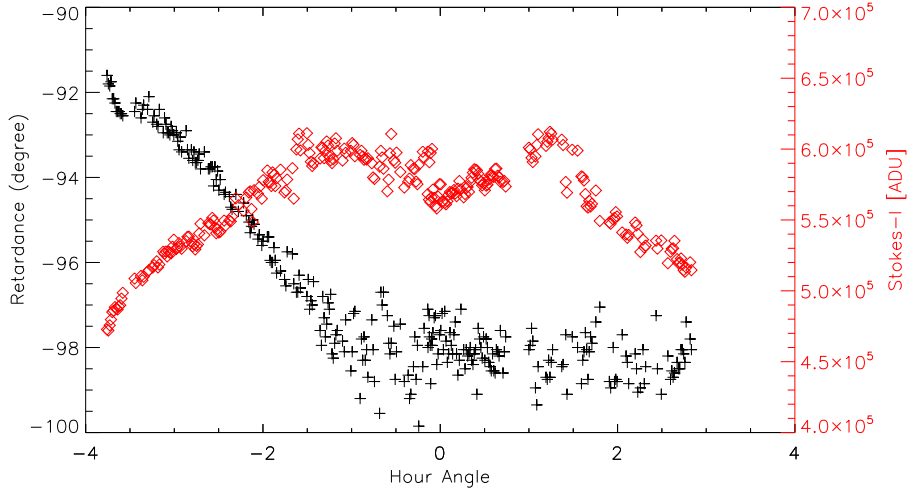}
\caption{Change of the retardance of the ICU retarder during a one-day continuous calibration (black-crosses) superposed on the variation of the Stokes $I$ after the demodulation (red-diamond).}
\label{Fig6}
\end{figure}

\subsection{Influence of Spectrograph azimuth angle}

To compensate the image rotation on the focus plane, the spectrograph, together with the beam-splitter ($M_{45}$), the polarization modulator and analyzer, needs to be synchronously rotated in the azimuth orientation ($\theta_{sp}$) during scientific observations. Since the Mueller matrix characterizes not only the optical properties but also the orientation of each element in the optical path, the response matrix $X(\theta_{sp} = 0)$ precisely measured at the original azimuth angle needs to be transformed to the $X(\theta_{sp} = \theta)$ by coordinate rotation in order to calibrate scientific observations carried out with the azimuth angle of $\theta$.

In the case of $\theta$ azimuth angle, the Eq.(4) describing the modulation process can be modified to be :

\begin{equation}
D \cdot I_{obs} = D \cdot M \cdot M_{M_{45}} \cdot Rot(\theta) \cdot M_T \cdot S_{sun}
\end{equation}
in which,  we define

\begin{equation}
\begin{split}
X(\theta_{sp} = \theta) = D \cdot M \cdot M_{M_{45}} \cdot Rot(\theta) \cdot M_{T} \\
\end{split}
\end{equation}
with the matrix of coordinate rotation

\begin{center}
\begin{equation}
Rot(\theta)=
\begin{bmatrix}
  1 & 0 &  0 &  0 \\
  0 & cos2\theta  &  sin2\theta &  0\\
  0 & -sin2\theta &  cos2\theta &  0 \\
  0 & 0 & 0 &  1
\end{bmatrix}
\end{equation}
\end{center}

 To realize the transformation form $X(\theta_{sp}=0)$ to $X(\theta_{sp}=\theta)$ according to Eq.(4) and Eq.(9), it is obliged  to have a good knowledge of the modulation matrix $M$ and the Mueller matrix of the beam-splitter $M_{M_{45}}$. To know the matrix $M$, as mentioned in Section 2, both the retardance of the modulator and the gear control precision of the modulator rotation are precisely measured in the laboratory. To know the matrix $M_{M_{45}}$, we firstly test it in the laboratory using the Mueller Matrix Measurement System, which is sensitive to $10^{-3}$ for each matrix element for weak polarization optical elements. The general principle of such measurement can be found in \cite{Ichimoto2006}. We also compare the $M_{M_{45}}$ with the result measured by using the NVST polarization observation system \citep{Peng2018}. A typical result of the $M_{M_{45}}$ is :

\begin{center}
\begin{equation}
M_{M_{45}}=
\begin{bmatrix}
  1.0000 & 0.0232 &  0.0007 &  0.0004 \\
  0.0227 & 1.0003 &  0.0023 &  -0.0115 \\
 -0.0004 & 0.0025 &  0.9006 &  0.4328 \\
  0.0008 & 0.0113 & -0.4360 &  0.8999
\end{bmatrix}
\end{equation}
\end{center}

One experiment of transformation is carried out on February 12th, 2019 as shown in Figure 7. We directly measure $X$ at the azimuth angle of $0^{o}$ and $30^{o}$, respectively, at different time. It is clearly seen that the response matrix dramatically changes in time due to the change of the azimuth angle. However as long as the parameters mentioned above are accurately given, the transformation from $X(\theta_{sp} = 30^{\circ})$ to $X(\theta_{sp} = 0^{\circ})$ can be correctly accomplished, i.e.,  the response matrix recovers the good continuity in time after transformation. In return, it is indicated that the measurement of Mueller matrices of optical elements and the response matrix itself is fairly precise.

\begin{figure}[htbp]
\centering
\includegraphics[width=12.cm,angle =0]{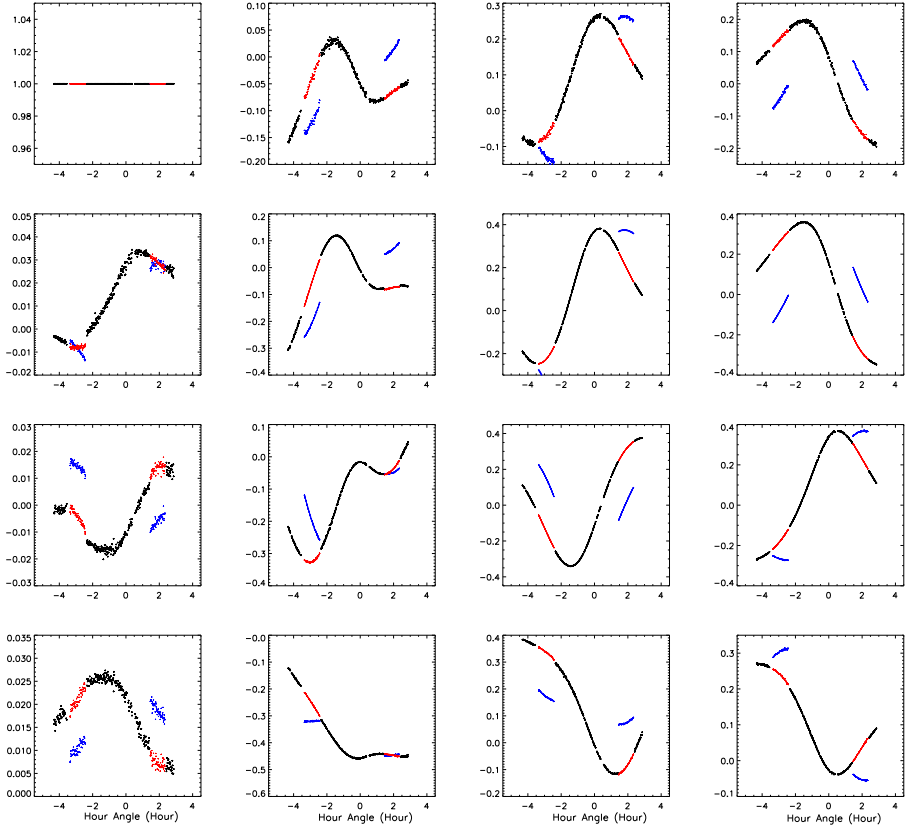}
\caption{ Response matrix $X$ of NVST as function of hour angle of the Sun on February 12th, 2019. Black dots: the response matrix measured with the azimuth angle of the spectrograph of $0^{o}$, i.e., $X(\theta_{sp}=0^{o})$. Blue dots: the response matrix measured with the azimuth angle of $30^{o}$,  $X(\theta_{sp}=30^{o})$. Red dots: the transformed response matrix from $\theta_{sp}=30^{o}$ to $\theta_{sp}=0^{o}$.}
\label{Fig7}
\end{figure}

\section{Application to Observations}
A $\delta$-class sunspot region (NOAA 12645, S10 W60) was observed in Fe I 532.4 nm using the NVST polarimeter on April 5th, 2017. The slit width is 0.1 mm corresponding to 0.45 arcsec on the solar image. Spatial sampling along the list is 0.082 arcsec pixel$^{-1}$ and the field-of-view is about 137 arcsec limited by the slit length.

We perform a 8-step polarization modulation and collect 20 frames at each modulation state. The exposure time of single image is 30 ms, and by collecting all frames at each modulation state, the typical noise level gets 10$^{-3}I_{c}$. As a result, it totally takes about 23 seconds to complete the modulation at one slit position. The azimuth angle of the spectrograph position is set at $\theta_{sp} = 0$ in both the calibration measurement and the scientific observation time, which happens to make the slit almost parallel to the solar radial orientation, going through the major sunspot and other satellite sunspots. Calibration measurements of the instrumental polarization are carried out before and after the scientific observation, respectively. Time interval of these two calibration is less than 30 minutes, but the response matrix apparently and linearly changes during this period. Therefore, we linearly interpolate the response matrix for the time of the scientific observation in order to calibrate the instrumental polarization at that time.

Before demodulating the observed $I_{obs}$, we applied the standard spectrum-data reduction routines, including dark current subtraction and flat-fielding.

\begin{figure}[htbp]
\centering
\includegraphics[width=12.cm,angle =0]{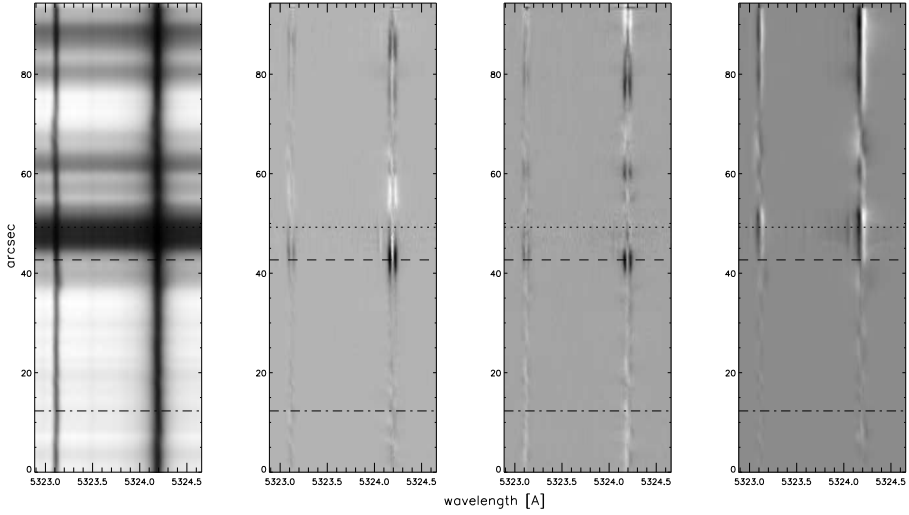}
\caption{2D Stokes spectra along one slit in  NOAA 12645.\textit{Left to right}: $I$, $Q/I$, $U/I$ and $V/I$. The dotted-, dashed- and dot-dashed-line indicate the typical sunspot umbra, penumbra and quiet sun areas, respectively, whose 1D Stokes profiles are plot in Figure 8 from top to bottom. }
\label{Fig8}
\end{figure}

\begin{figure}[htbp]
\centering
\includegraphics[width=12.cm,angle =0]{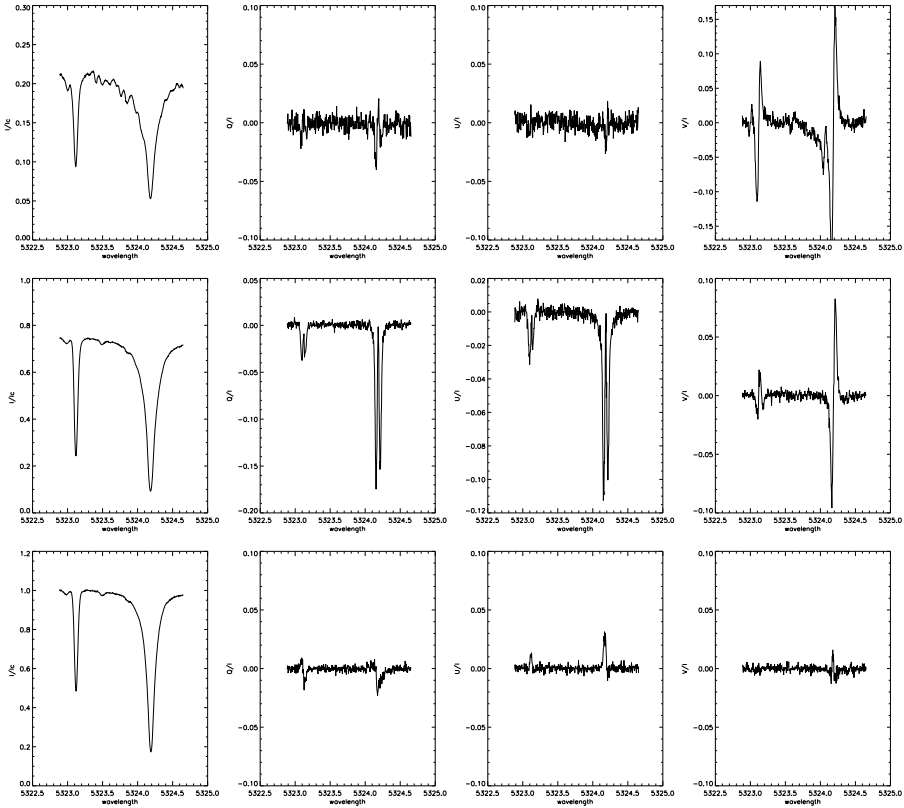}
\caption{Stokes profiles observed in NOAA 12645. \textit{Left to right}: $I/I_{c}$, $Q/I$, $U/I$ and $V/I$. $I_{c}$ is the continuum intensity of the quiet Sun profile. \textit{Top to bottom}: Stokes profiles taken from the sunspot umbra, penumbra and quiet Sun as marked by different lines in Figure8.}
\label{Fig9}
\end{figure}

The Stokes vectors $S^{sun}_{\lambda}$ are derived after calibrating instrumental polarization. However there still exists some residual cross-talk between different Stokes vectors even after the calibration. The relationship to the real Stokes vectors from the Sun, $S^{real}_{\lambda}$, can be expressed as the following Eq.(12) with an assumption that the most serious cross-talk comes from Stokes $I$.

\begin{equation}
\begin{split}
S^{sun}_{\lambda} = S^{real}_{\lambda} + a_{_{I\rightarrow S}}I_{\lambda} \\
\frac{S^{sun}_{\lambda}}{I_{\lambda}} = \frac{S^{real}_{\lambda}}{I_{\lambda}} + a_{_{I\rightarrow S}}
\end{split}
\end{equation}

Here $a_{I\rightarrow S}$ means the proportion of the cross-talk coming from the Stokes $I$ and it is wavelength- and spatial-independent. If we take an expression of $S_{\lambda}/I_{\lambda}$, we can easily estimate this cross-talk in the continuum wavelength from the 2D Stokes image, since the local polarization signal $S^{real}$ is supposed to be zero. From the observation on NOAA 12645, we estimate that the $a_{I\rightarrow q}$, $a_{I\rightarrow u}$ and $a_{I \rightarrow v}$ are $6.5\times10^{-3}$, $1.2\times10^{-3}$ and $5.5\times10^{-3}$, respectively. The averaged value is about 4$\times 10^{-3}$.

After this estimation, we force the polarization signal at the continuum to be zero and show the resulted 2D Stokes images in Figure 8. Spatial changes along the slit are clearly exhibited in full Stokes profiles. A large scale Evershed flow around the main sunspot is clearly revealed in Stokes $I$,  which is demonstrated as an obvious tilt along the slit direction of the spectrum inside the sunspot. Based on the Doppler shift calculation, it is found that the line-of-sight speed varies from around 1 km/s at the border between the umbra and the penumbra to a maximum of 1.8 km/s in the middle of the penumbra and falls off to zero at the edge of the penumbra. In addition, typical symmetric or anti-symmetric Stokes $Q$, $U$ and $V$ profiles are present around sunspot. The full Stokes profiles taken from the umbra, penumbra and quiet region are shown in Figure 9. From these observed results, we can estimate the polarimetric sensitivity, i.e., the standard deviation of the noise level in continuum. In Figure 9, the noise level in Stokes $V/I$ is about $2.1 \times 10^{-3}$ and the results of Stokes $Q/I$ and $U/I$ are very close to it. In fact, the noise level is closely related to the total integration time (i.e., the number of integrated frames taken at one modulation state). Some investigations with different integration time are listed in Table 2. The noise level increases with the integration time decreases, and it reaches 3.3$\times 10^{-3}$ in 14 sec and 5.2$\times 10^{-3}$ in 10 sec.

\begin{table}
\begin{center}
\caption[]{Noise level in continuum with different integration time.}\label{Tab:2}
{\begin{tabular}{cccc}
\hline
 Integration time (s)  & 23   & 14  & 10  \\
\hline
 Integrated Frame number (n)  & 20  &  10 & 5 \\
\hline
  noise level  & 2.1$\times 10^{-3}$  &  3.3$\times 10^{-3}$ & 5.2$\times 10^{-3}$ \\

\hline

\end{tabular}}
\end{center}
\end{table}

\section{Discussion and Summary}

Spectro-polarimetric observation is the third phase of the NVST multi-wavelength spectrograph operation. Initially, spectrograph installation and its salient function were accomplished in 2013 \citep{WangR2013}. It is followed by the precise data reduction and 2D field scan \citep{Cai2018}. The efforts presented in this paper is to develop the capability of spectro-polarimetric observation for future vector magnetic field measurements. As mentioned above, these efforts are only the preliminary steps in view of the high precision polarimetry observation, particularly, taking into account the slow modulation rate, the 2D spatial scan and the combination with the AO system, etc.

In this work, we mainly deal with important topics of the polarimetric calibration of the instrument. The response matrix of the NVST can be accurately determined by utilizing ICU. It is verified that both the facilities and measurement procedures are fairly stable and credible, as shown by the comparison of the calibration results on February 16th, 2017 (Figure 4) and on February 12th, 2019 (Figure 7). The time variation of each matrix element shows almost same morphology and magnitude, which suggests that the 2-year gap has no obvious effect on the daily evolution of the response matrix from the date of 12th to 16th.


The calibration accuracy can reach up to 3$\pm$ 1.3$\times10^{-3}$ with the consideration that the retardance of the ICU retarder ($\delta^{c}_{r}$) is a free variable. Variation of the retardance and the intensity of Stokes $I$ after the demodulation show a good correlation during the continuous calibration. The straightforward reason is not clear, but the temperature increase during the continuous observation at the secondary focus where ICU is located may be a candidate to explain this feature. More detection and experiments are needed to confirm it. In addition, from the comparison of Figure 5 with Figure 6, we find that the large fluctuation of the calibration residual is present when this retardance is calculated with a relatively large scattering, which indicates that the main contribution to the calibration errors comes from the imprecise measurement of this retardance.

Based on the spectro-polarimetric observation for an active region NOAA 12645, it is verified that, firstly, the calibration result is credible and the instrumental polarization is removed largely. Secondly, after calibrating the instrumental polarization, the residual cross-talk from Stokes $I$ to Stokes $Q, U$ and $V$ is about 4$\times 10^{-3}$ on average, which is consistent with the calibration accuracy and close to the photon noise (or statistical noise). However we do not investigate the residual cross-talk among Stokes $Q, U$ and $V$ for the time being. We intend to perform a statistic analysis for amount of Stokes profiles taken from isolated $\alpha$-class sunspots when they are approaching to the disk center and investigate the correlation among the Stokes profile shapes \citep{Kuhn1994}. Thirdly, it is revealed that an integration time over 20 seconds is needed for present system in order to archive a polarimetric sensitivity or a detection limit of $10^{-3}$. Slow rate of the modulation is indeed an issue for our present system and a continuous modulation process is now under consideration.

As mentioned in Figure 1, both the field scanner and AO system have not been included into the present optical system yet. Up to now, we have carried out the Mueller matrix measurement for each elements independently in the laboratory. We will integrate these two elements in the upcoming instrument update.

An IDL-based software package has been developed for the spectrum data reduction and polarimetric calibration. Further more, we have accomplished the full Stokes inversion using the Helix code \citep{Lagg2004}, which is based on the Unno-Rachkowsky analytic solution of the radiative transfer for polarized radiation in a Milne-Eddington atmosphere approximation. The vector magnetic field and other physical quantities are retrieved. However considering the main issue of this paper, we do not show the inversion results in this paper.

\begin{acknowledgements}
This work was supported by the National Natural Science Foundation of China under grants 11873091, 11773040, 11773072 and 11373044. We are grateful to the anonymous reference for his thoughtful discussion and help clarifying this manuscript.
\end{acknowledgements}


\end{document}